\documentclass[12pt]{article}
\usepackage{amssymb,amsmath}
\usepackage[noblocks]{authblk}
\usepackage[top=0.75in, bottom=0.75in, left=0.75in, right=0.75in, dvips]{geometry}
\usepackage{caption}
\begin{document}
\textwidth 10.0in
\textheight 9.0in
\topmargin -0.60in
\title{Non-Trivial Ghosts and Second Class Constraints}
\author[1]{Farrukh Chishtie}
\author[1,2]{D.G.C. McKeon}
\affil[1] {Department of Applied Mathematics, The
University of Western Ontario, London, ON N6A 5B7, Canada}
\affil[2] {Department of Mathematics and
Computer Science, Algoma University, Sault St.Marie, ON P6A
2G4, Canada}

\maketitle

\maketitle

\begin{abstract}
In a model in which a vector gauge field $W_\mu^a$ is coupled to an antisymmetric tensor field $\phi_{\mu\nu}^a$ possessing
a pseudoscalar mass, it has been shown that all physical degrees of freedom reside in the vector field.  Upon quantizing this model
using the Faddeev-Popov procedure, explicit calculation of
the two-point functions $<\phi\phi >$ and $<W \phi>$ at one-loop order seems to have yielded the puzzling result that the effective
action generated by radiative effects has more physical degrees of freedom than the original classical action. In this paper we point
out that this is not in fact a real effect, but rather appears to be a consequence of having ignored a ``ghost''
field arising from the contribution
to the measure in the path integral arising from the presence of non-trivial second-class constraints. These ghost fields couple to
the fields $W_\mu^a$ and $\phi_{\mu\nu}^a$, which makes them distinct from other models involving ghosts arising from second-class
constraints (such as massive Yang-Mills (YM) models) that have been considered, as in these other models such ghosts decouple.  As
an alternative to dealing with second class constraints, we consider introducing a ``Stueckelberg field'' to eliminate second-class
constraints in favour of first-class constraints and examine if it is possible to then use the Faddeev-Popov quantization procedure.
In the Proca model, introduction of the Stueckelberg vector is equivalent to the Batalin-Fradkin-Tyutin (BFT) approach to converting
second-class constraints to being first class through the introduction of new variables. However, introduction of a Stueckelberg
vector is not equivalent to the BFT approach for the vector-tensor model.  In an appendix, the BFT procedure is applied to the pure tensor
model and a novel gauge invariance is found. In addition, we also consider extending the Hamiltonian so that half of the
second-class constraints become first-class and the other half become associated gauge conditions. We also find for this tensor-vector theory that when converting the phase space path integral to the configuration space path integral, a non-trivial contribution to the measure arises that is not manifestly covariant and which is not simply due to the presence of second class constraints.

\end{abstract}

\section{Introduction}
Some time ago, a model in which a non-Abelian vector gauge field coupled
to an antisymmetric tensor field that has a pseudo-scalar mass term was introduced
{[}1{]}. The original motivation for considering this model was to
see if the mass parameters occurring in this model could induce a
pole in the propagator for the vector field that would be away from
the massless limit, thereby providing an alternative to the Higgs
mechanism for giving a mass to vector fields. Although this hope
was not realized, it became apparent that this model is interesting
for an unexpected reason: the presence of a pseudo-scalar mass term
for the antisymmetric tensor field serves to eliminate all physical degrees of freedom
from the model except for the usual transverse degrees of freedom
present in the vector gauge field. This elimination of any physical
degrees of freedom associated with the tensor field is surprising,
notably because of the highly non-trivial way it occurs in the original
action, but also because normally introduction of a mass term into
a gauge invariant model (such as Yang-Mills theory) serves to increase
the number of physical degrees of freedom, not decrease their number.
The model we consider has the classical Lagrangian {[}1{]}

%eq1
\begin{align}
\mathcal{L}_{c1} & =-\frac{1}{4}F_{\mu\nu}^{a}F^{a\mu\nu}+\frac{1}{12}G_{\mu\nu\lambda}^{a}G^{a\mu\nu\lambda}+\frac{m}{4}\epsilon^{\mu\nu\lambda\sigma}\phi_{\mu\nu}^{a}F_{\lambda\sigma}^{a}\\
 & \qquad+\frac{\mu^{2}}{8}\epsilon^{\mu\nu\lambda\sigma}\phi_{\mu\nu}^{a}\phi_{\lambda\sigma}^{a}\nonumber
\end{align}
with $m$ and $\mu$ being mass parameters and %eq2
\begin{equation}
F_{\mu\nu}^{a}=\partial_{\mu}W_{\nu}^{a}-\partial_{\nu}W_{\mu}^{a}+f^{abc}W_{\mu}^{b}W_{\nu}^{c}
\end{equation}
%eq3
\begin{equation}
G_{\mu\nu\lambda}^{a}=D_{\mu}^{ab}\phi_{\nu\lambda}^{b}+D_{\nu}^{ab}\phi_{\lambda\mu}^{b}+D_{\lambda}^{ab}\phi_{\mu\nu}^{b}
\end{equation}
and %eq4
\begin{equation}
\hspace{-0.5cm}D_{\mu}^{ab}=\partial_{\mu}\delta^{ab}+f^{apb}W_{\mu}^{p}\quad([D_{\mu},D_{\nu}]^{ab}=f^{apb}F_{\mu\nu}^{p})
\end{equation}
$(\eta_{\mu\nu}={\rm {diag}(+---),\epsilon^{0123}=+1)}$.

In ref. {[}1{]}, a constraint analysis {[}5-13{]} of this model shows
that with twenty initial degrees of freedom (dof) in phase space ($\phi_{\mu\nu}^{a}-6\mathrm{dof}$,
$W_{\mu}^{a}-4\mathrm{dof}$, plus associated momenta) and five first-class
constraints, five gauge conditions and six-second class constraints,
there are four net physical degrees of freedom, provided $\mu^{2}\neq0$.
In the Abelian limit, explicit elimination of fields by use of equations
of motion that are free of time derivatives shows that these are the
usual transverse polarizations of the vector $W_{\mu}$ and their
associated momenta.

It was clearly of interest to see if the interactions between
the tensor and vector fields appearing in eq. (1) would somehow include
dynamical degrees of freedom when radiative effects were taken into
account. The one-loop contributions to the two-point functions $<W_{\mu}^{a}W_{\nu}^{b}>$
{[}2{]}, $<\phi_{\mu\nu}^{a}\phi_{\lambda\sigma}^{b}>$ {[}3{]} and
$<W_{\mu}^{a}\phi_{\lambda\sigma}^{b}>$ {[}4{]} were all computed
using the Feynman rules derived from the usual Fadeev-Popov quantization
procedure that works so well for Yang-Mills theory {[}17{]}. In this
approach to quantization, the usual gauge invariance whose infinitesimal
form is %eq5
\begin{equation}
\delta W_{\mu}^{a}=D_{\mu}^{ab}\theta^{b}
\end{equation}
%eq6
\begin{equation}
\hspace{0.5cm}\delta\phi_{\mu\nu}^{a}=f^{abc}\phi_{\mu\nu}^{b}\theta^{c}
\end{equation}
results in the necessity of gauge fixing. The Feynman gauge fixing
Lagrangian %eq7
\begin{equation}
\mathcal{L}_{gf}=-\frac{1}{2}(\partial\cdot W^{a})^{2}
\end{equation}
leads normally to the introduction of the usual Faddeev-Popov (FP)
ghost fields $c^{a}$, $\overline{c}^{a}$ whose action %eq8
\begin{equation}
\mathcal{L}_{gh}=\overline{c}^{a}(\partial\cdot D^{ab})c^{b}
\end{equation}
involves a coupling of the ghosts to the vector. The effective action
$\mathcal{L}_{c1}+\mathcal{L}_{gf}+\mathcal{L}_{gh}$ (eqs. (1,7,8))
has the usual BRST invariance {[}6-{]}.

Using the Feynman rules based on $\mathcal{L}_{c1}+\mathcal{L}_{gf}+\mathcal{L}_{gh}$,
an explicitly calculation of the one-loop two point function $<W_{\mu}^{a}W_{\nu}^{b}>$
results in a complete cancellation of all diagrams involving the field
$\phi_{\mu\nu}^{a}$ {[}2{]}; the result is identical to what arises
from a pure YM theory. This is not unexpected as the analysis of the
canonical structure of the classical theory indicates that all degrees
of freedom in the model reside solely in the vector field.

However, an analogous calculation of the one-loop, two-point functions
$<\phi_{\mu\nu}^{a}\phi_{\lambda\sigma}^{b}>$ and\linebreak{}
$<\phi_{\mu\nu}^{a}W_{\lambda}^{b}>$ using dimensional regularization
show that these diverge {[}3,4{]}. Only by using non-local counter
terms proportional to $m$ can these divergences be removed, implying
that $\phi_{\mu\nu}^{a}$ develops degrees of freedom radiatively.
This is most peculiar, especially since the shift in the antisymmetric tensor field
so that %eq9
\begin{equation}
\phi_{\mu\nu}^{a}=\chi_{\mu\nu}^{a}-\frac{m}{\mu^{2}}F_{\mu\nu}^{a}
\end{equation}
leads to %eq10
\begin{align}
\mathcal{L}_{c1} & =-\frac{1}{4}F_{\mu\nu}^{a}F^{a\mu\nu}+\frac{1}{12}H_{\mu\nu\lambda}^{a}H^{a\mu\nu\lambda}+\frac{\mu^{2}}{8}\epsilon^{\mu\nu\lambda\sigma}\chi_{\mu\nu}^{a}\chi_{\lambda\sigma}^{a}\nonumber \\
 & \qquad\qquad+\frac{m^{2}}{8\mu^{2}}\epsilon^{\mu\nu\lambda\sigma}F_{\mu\nu}^{a}F_{\lambda\sigma}^{a}.
\end{align}
$(H_{\mu\nu\lambda}^{a}\equiv D_{\mu}^{ab}\chi_{\nu\lambda}^{b}+\ldots)$\\
 The last term in eq. (10) is topological and does not contribute
to perturbative calculations; the remaining terms are identical in
form to those in eq. (1) with $m=0$. Since by refs. {[}3,4{]} $<\phi\phi>$
and $<\phi W>$ appear to have divergent parts proportional to $m$,
this would mean that $<\chi\chi>$ and $<\chi W>$ should both be
free of divergences. However, this would all be apparently inconsistent
with eq. (9), as this equation superficially implies that %eq11
\begin{equation}
<\phi\phi>=<\chi\chi>-\frac{m}{\mu^{2}}(<\chi F>+<F\chi>)+\frac{m^{2}}{\mu^{4}}<FF>
\end{equation}
as well as %eq12
\begin{equation}
<\chi\chi>=<\phi\phi>-\frac{m}{\mu^{2}}(<\phi F>+<F\phi>)+\frac{m^{2}}{\mu^{4}}<FF>.
\end{equation}
The results of refs. {[}3,4{]} are inconsistent with eqs. (11,12).

These inconsistencies have motivated us to see if the structure of
the model itself somehow invalidates the quantization procedure used
to compute radiative effects in refs. {[}2-4{]}. It turns out that
the presence of second-class constraints in the model leads to a non-trivial
(field dependent) contribution to the measure of the path integral
that is not taken into account when using the Fadeev-Popov procedure
{[}17{]} to ``factor
 out the superfluous degrees of
freedom associated with gauge invariance of eqs. (5,6). Although the
contribution of second-class constraints to the measure of the path
integral have been understood for some time {[}21, 22{]}, the model
of eq. (1) provides the first example we know of in which the second-class
constraints make a non-trivial (viz. field-dependent) contribution
to the measure, thereby necessitating thereby necessitating the introduction
of a new type of Grassmann ``ghost" field to the effective
action. In the following section we explicitly show how this new ghost
arises. Following that, in section 3 we convert the path integral
from phase space to configuration space using the approach of ref.
{[}24{]}. In contrast to what happens in Yang-Mills theory {[}20,
43{]}, the resulting path integral in configuration space is not manifestly
covariant.

\section{Second-Class Constraints}

In attempting to compute radiative effects in YM theory using the
same approach that worked in quantum electrodynamics, Feynman encountered
an inconsistency that could be overcome by introducing Fermionic scalar
``ghost'' fields {[}14{]}. These FP fields were introduced more
formally by Mandelstam {[}15{]}, DeWitt {[}16{]} and Faddeev-Popov
{[}17{]}. (For a more general discussion of gauge fixing which leads
to having two Fermionic and one Bosonic ghost, see refs. {[}18,19{]}.)
This approach has been adequate for dealing with YM gauge theories
coupled to matter (as is appropriate for the standard model) but a
more general discussion is needed to eliminate superfluous degrees
of freedom in more complicated models.

Beginning with the canonical analysis of constrained systems, Faddeev
examined how a system with only first-class constraints could be quantized
using the path integral {[}20{]}. He argued that for YM theory this
approach is equivalent to that of ref. {[}17{]} in which ``gauge
orbits'' are factored out of the path integral over all field configurations.
His analysis was extended to systems with second-class constraints
by Fradkin {[}21{]} and Senjanovic {[}22{]}. In a system with a denumerable
number of degrees of freedom $(q^{i}(t),p_{i}(t))$ in phase space,
the matrix element for the $S$-matrix is %eq13
\begin{equation}
<{\mathrm{out}}|S|{\mathrm{in}}>=\int\exp i\int_{-\infty}^{\infty}dt[p_{i}\dot{q}^{i}-H(q^{i},p_{i})]D\mu(q^{i}(t),p_{i}(t))
\end{equation}
where the measure is %eq14
\begin{align}
D\mu(q^{i}(t),p_{i}(t)) & =\prod_{a}\delta(\phi_{a})\delta(\chi_{a})\det\left\lbrace \phi_{a},\chi_{b}\right\rbrace \\
 & \qquad\qquad\prod_{b}\delta(\theta_{b})\mathrm{det}^{1/2}\left\lbrace \theta_{a},\theta_{b}\right\rbrace Dq^{i}(t)Dp_{i}(t)\nonumber
\end{align}
where $q^{i}(t)\rightarrow(q_{{\rm {out}}}^{i},q_{{\rm {in}}}^{i})$
as $t\rightarrow\pm\infty$, $H$ is the canonical Hamiltonian, $\phi_{a}$,
$\chi_{a}$ and $\theta_{a}$ are the first-class constraints, associated
gauge conditions and second-class constraints respectively, and $\left\lbrace ,\right\rbrace $
denotes the Poisson Bracket (PB). Eq. (13) is independent of the choice
of gauge conditions $\chi_{a}$ provided $\det\left\lbrace \phi_{a},\chi_{b}\right\rbrace \neq0$.
Formally the factor of $\Theta=\det^{1/2}\left\lbrace \theta_{a},\theta_{b}\right\rbrace $
appears in such theories as YM theories with a Proca mass, scalar
theories quantized using light-cone coordinates and models with magnetic
monopoles {[}22{]}; it is completely absent in massless YM theories
{[}23{]}. However, there do not appear to be any examples in the literature
for field theoretic models in which $\Theta$ explicitly involves
either $q^{i}(t)$ or $p_{i}(t)$ and hence $\Theta$ can generally
be absorbed into a normalization factor.

We note that it is possible to absorb the functional integration over
the momentum $p_{i}(t)$ in eq. (13) entirely into the measure, leaving
the argument of the exponential to be $i\int_{-\infty}^{\infty}dtL(q_{i},\dot{q}^{i})$
where $L$ is the Lagrangian of the system {[}24{]}. It is also possible
to generalize the quantization procedure to deal with gauge theories
possessing open algebras. (For reviews, see refs. {[}7-13{]}.)

It has also been demonstrated that second-class constraints can be
converted into first-class ones by the introduction of new variables
accompanied by an extension of the Hamiltonian so that the modified
theory possesses a gauge invariance not originally present {[}25-28{]}.
We will illustrate how this ``BFT'' procedure works by considering
a massive Proca vector field whose action is %eq15
\begin{equation}
\mathcal{L}_{p}=-\frac{1}{4}(\partial_{\mu}A_{\nu}-\partial_{\nu}A_{\mu})^{2}+\frac{m^{2}}{2}A_{\mu}A^{\mu}.
\end{equation}
If $\pi^{0}$ and $\pi^{i}$ are the momenta associated with $A_{0}$
and $A_{i}$ respectively, it is easy to see that there is a primary
second-class constraint %eq16
\begin{equation}
\theta_{1}=\pi^{0}=0
\end{equation}
and a secondary second-class constraint %eq17
\begin{equation}
\theta_{2}=\partial_{i}\pi^{i}+m^{2}A_{0}=0
\end{equation}
along with the canonical Hamiltonian %eq18
\begin{equation}
\mathcal{H}_{c}=\frac{1}{2}\pi^{i}\pi^{i}+\frac{1}{4}F_{ij}F_{ij}-\frac{m^{2}}{2}(A_{0}^{2}-A_{i}A_{i})-A_{0}\partial_{i}\pi^{i}.
\end{equation}
There are no first-class constraints, which is consistent with there
being no gauge invariance in the Proca Lagrangian.

With the BFT procedure, we can introduce fields $\eta_{1}$ and $\eta_{2}$
such that {[}29,30{]} %eq19
\begin{equation}
\left\lbrace \eta_{i},\eta_{j}\right\rbrace =m^{2}\epsilon_{ij}
\end{equation}
and then replace eqs. (16-18) by {[}28{]} %eq20
\begin{equation}
\hspace{-2cm}\overline{\theta}_{1}=\pi^{0}+\eta_{1}
\end{equation}
%eq21
\begin{equation}
\overline{\theta}_{2}=\partial_{i}\pi^{i}+m^{2}A_{0}+\eta_{2}
\end{equation}
%eq22
\begin{equation}
\overline{\mathcal{H}}_{p}=\mathcal{H}_{p}-(\partial_{i}A_{i})\eta_{1}-\frac{1}{m^{2}}(\partial_{i}\pi^{i}+m^{2}A_{0})\eta_{2}-\frac{1}{2m^{2}}(\eta_{2}^{2}+\eta_{1}\nabla^{2}\eta_{1}).
\end{equation}
We now find that $\overline{\theta}_{1}$ and $\overline{\theta}_{2}$
are first-class constraints that have a weakly vanishing PB with $\overline{\mathcal{H}}_{p}$.
A gauge invariance that consequently occurs in this modified theory
can be worked out using either the approach of ref. {[}31{]} or ref.
{[}32{]}. No second class constraint consequently occurs in the path
integral of eqs. (13,14) which are associated with this model. In
the gauge $\eta_{1}=\eta_{2}=0$ the Proca model is recovered.

The introduction of extra fields to preserve a gauge invariance was
an idea that originated with Stueckelberg {[}33{]}, though not in
the context of the constraint formalism. If the action of eq. (15)
were replace by %eq23
\begin{equation}
\mathcal{L}_{s}=-\frac{1}{4}F_{\mu\nu}F^{\mu\nu}+\frac{m^{2}}{2}(A_{\mu}+\frac{1}{m}\partial_{\mu}\sigma)^{2}
\end{equation}
where $\sigma$ is the ``Stueckelberg scalar'', then one has the
gauge invariance %eq24
\begin{equation}
\delta A_{\mu}=\partial_{\mu}w
\end{equation}
%eq25
\begin{equation}
\hspace{0.5cm}\delta\sigma=-mw.
\end{equation}
One can show that the introduction of the Stueckelberg scalar into
the Proca Lagranian is equivalent to the introduction of the fields
$\eta_{1}$ and $\eta_{2}$ through the BFT procedure by performing
a constraint analysis of $\overline{\mathcal{L}}_{s}=\mathcal{L}_{s}-m\partial_{0}(\sigma A_{0})$.
From $\overline{\mathcal{L}}_{s}$ we find only the primary and secondary
first-class constraints %eq26
\begin{equation}
\hspace{-2cm}\pi^{0}+m\sigma=0
\end{equation}
%eq27
\begin{equation}
\partial_{i}\pi^{i}+m^{2}A_{0}+m\pi=0
\end{equation}
which suggests, upon comparing eqs. (20,21) with eqs. (25,26), that
%eq28
\begin{equation}
\eta_{1}=m\sigma
\end{equation}
%eq29
\begin{equation}
\eta_{2}=m\pi.
\end{equation}
(There are no second-class or tertiary constraints.)\\
 The Hamiltonian that follows from $\overline{\mathcal{L}}_{s}$ is
$\mathcal{H}_{p}$ in eq. (22) provided one employs the constraint
of eq. (21). (The fields $A_{0}$, $A_{i}$ and $\sigma$ have conjugate
momenta $\pi^{0}$, $\pi^{i}$ and $\pi$ respectively.)

The Stueckelberg model of eq. (23) can be quantized using the FP procedure.
using the gauge fixing Lagrangian %eq30
\begin{equation}
\mathcal{L}_{gf}=\frac{-1}{2\alpha}(\partial_{\mu}A^{\mu}-\alpha m\sigma)^{2}
\end{equation}
is particularly convenient as in this gauge $A_{\mu}$ and $\sigma$
decouple and the renormalizability of a model in which $A_{\mu}$
is coupled to a conserved current becomes apparent. In the gauge $\sigma=0$,
we recover the Proca model and one should in principle consider how
the second-class constraints of eqs. (16,17) contribute to the measure
of the classical action, though in practice this measure can be ignored
as $\left\lbrace \theta_{1},\theta_{2}\right\rbrace =-m^{2}$ which
is just a constant.

The constraints of the model in eq. (1) are much more complicated
than those of the Proca model; in particular the PB of the second
class constraints is no longer field independent. Following the procedure
used in ref. {[}1{]} to analyze the constraints in the Abelian limit
of $\mathcal{L}_{c1}$, we define %eq31a-d
\[
U^{a}=W_{0}^{a},\qquad V_{i}^{a}=W_{i}^{a},\qquad A_{k}^{a}=\phi_{0k}^{a},\qquad B_{k}^{a}=\frac{1}{2}\epsilon_{k\ell m}\phi_{\ell m}^{a}\,.\eqno(31a-d)
\]
When $\mathcal{L}_{c1}$ is written in terms of these fields so that
%eq nonumber
\[
\hspace{-4cm}\mathcal{L}=-\frac{1}{4}(F_{ij}^{a})^{2}+\frac{1}{2}(\dot{V}_{i}^{a}-D_{i}^{ab}U^{b})^{2}+\frac{1}{2}(\dot{B}_{i}^{a}+f^{abc}U^{b}B_{i}^{c})^{2}
\]
\[
-(\dot{B}_{i}^{a}+f^{amn}U^{m}B_{i}^{n})\epsilon_{ijk}(D_{j}A_{k})^{a}+\frac{1}{2}(\epsilon_{ijk}(D_{j}A_{k})^{a})^{2}-\frac{1}{2}(D_{i}B_{j})^{a}(D_{j}B_{i})^{a}
\]
\[
+\mu^{2}A_{i}^{a}B_{i}^{a}+\frac{m}{2}\left[\epsilon_{ijk}A_{i}^{a}F_{jk}^{a}+2B_{k}^{a}(\dot{V}_{k}^{a}-\partial_{k}U^{a}+f^{abc}U^{b}V_{k}^{c})\right]
\]
we find that their respective canonical momenta are %eq32a-d
\[
\pi^{Ua}=0,\quad\pi_{i}^{Va}=\partial_{0}V_{i}^{a}-(D_{i}U)^{a}+mB_{i}^{a},\quad\pi_{i}^{Aa}=0,\quad\pi_{i}^{Ba}=(D_{0}B_{i})^{a}-\epsilon_{ijk}(D_{j}A_{k})^{a},\eqno(32a-d)
\]
from which follows the canonical Hamiltonian %eq33
\[
\mathcal{H}_{c}=\frac{1}{2}\pi_{i}^{Va}\pi_{i}^{Va}+\frac{1}{2}\pi_{i}^{Ba}\pi_{i}^{Ba}+\pi_{i}^{Va}(D_{i}U)^{a}+\pi_{i}^{Ba}\epsilon_{ijk}(D_{j}A_{k})^{a}
\]
\[
\qquad+f^{abc}U^{a}\pi_{i}^{Bb}B_{i}^{c}+\frac{1}{4}F_{ij}^{a}F_{ij}^{a}+\frac{1}{2}(D_{i}B_{j})^{a}(D_{j}B_{i})^{a}
\]
\[
\qquad\qquad-\mu^{2}A_{i}^{a}B_{i}^{a}-m\pi_{i}^{Va}B_{i}^{a}-\frac{m}{2}\epsilon_{ijk}A_{i}^{a}F_{jk}^{a}+\frac{m^{2}}{2}B_{i}^{a}B_{i}^{a}.\eqno(33)
\]
The primary constraints of eqs. (32a,c) lead respectively to the secondary
constraints %eq34a
\[
S^{a}=(D_{i}\pi_{i}^{V})^{a}+f^{abc}B_{i}^{b}\pi_{i}^{Bc}\eqno(34a)
\]
%eq34b
\[
\hspace{-0.5cm}S_{i}^{a}=\epsilon_{ijk}D_{j}^{ab}\pi_{k}^{Bb}-\mu^{2}B_{i}^{a}-\frac{m}{2}\epsilon_{ijk}F_{jk}^{a}.\eqno(34b)
\]
These constraints have the PB algebra %eq35a-c
\[
\left\lbrace S_{i}^{a},S_{j}^{b}\right\rbrace =0,\quad\left\lbrace S^{a},S^{b}\right\rbrace =f^{abc}S^{c},\quad\left\lbrace S_{i}^{a},S^{b}\right\rbrace =f^{abc}S_{i}^{c}.\eqno(35a-c)
\]

The PB of $S^{a}$ with $\int\mathcal{H}_{c}dx$ weakly vanishes.
However, the PB of $S_{i}^{a}$ with $\int\mathcal{H}_{c}dx$ yields
the tertiary constraint %eq36
\[
T_{i}^{a}=-\mu^{2}\pi_{i}^{Ba}+\mu^{2}f^{abc}U^{b}B_{i}^{c}+\epsilon_{ijk}\left[f^{abc}\left(\pi_{j}^{Vb}\pi_{k}^{Bc}\right.\right.
\]
\[
\left.+(D_{j}U)^{b}\pi_{k}^{Bc}-mB_{j}^{b}\pi_{k}^{Bc}\right)-D_{j}^{ab}(f^{bcd}U^{c}\pi_{k}^{Bd})
\]
\[
\left.+(D_{j}D_{\ell}D_{k}B_{\ell})^{a}-m(D_{j}D_{k}U)^{a}\right].\eqno(36)
\]

We see that since %eq37a
\[
\hspace{-5cm}\left\lbrace T_{i}^{a},S^{b}\right\rbrace =f^{abc}T_{i}^{c}\eqno(37a)
\]
%eq37b
\[
\left\lbrace S_{i}^{a},T_{\ell}^{x}\right\rbrace =f^{apm}f^{xpn}\left(\delta_{i\ell}\pi_{k}^{Bm}\pi_{k}^{Bn}-\pi_{\ell}^{Bm}\pi_{i}^{Bn}\right)
\]
\[
+\delta_{i\ell}(D_{j}D_{k}D_{k}D_{j}-D_{j}D_{j}D_{k}D_{k})^{ax}
\]
\[
+(D_{\ell}D_{i}D_{j}D_{j}+D_{j}D_{j}D_{\ell}D_{i}-D_{j}D_{i}D_{\ell}D_{j}-D_{\ell}D_{j}D_{j}D_{i})^{ax}
\]
\[
+\mu^{2}\left[\mu^{2}\delta^{ax}\delta_{i\ell}+\epsilon_{i\ell m}\left(-D_{m}^{ab}f^{xbz}U^{z}\right.\right.\eqno(37b)
\]
\[
\left.\left.-f^{axp}\pi_{m}^{Vp}-f^{axp}(D_{m}U)^{p}+(D_{m}^{xy}f^{ayw}U^{w})+mf^{axy}B_{m}^{y}\right)\right]
\]
both $S_{i}^{a}$ and $T_{i}^{a}$ are irreducible second class even
in the Abelian limit if $\mu^{2}\neq0$. There are five first-class
constraints $(\pi^{Ua},\pi_{i}^{Aa},S^{a})$ and six second-class
constraints $(S_{i}^{a},T_{i}^{a})$ which accounts for why sixteen
of the twenty degrees of freedom in phase space ($W_{\mu}^{a},\phi_{\mu\nu}^{a}$
and their associated momenta) are non-physical. The remaining four
degrees of freedom are the two transverse polarizations of $W_{\mu}^{a}$
and their conjugate momenta. The gauge transformation generated by
$(\pi^{Ua},S^{a})$ is the usual non-Abelian gauge transformation
of eqs. (5,6) which is a closed, irreducible gauge transformation
that forms a Lie algebra {[}13{]}. Just as in the canonical analysis
of the first order Einstein-Hilbert action in $d>2$ dimensions, the
second class constraints must be eliminated before the algebra of
first class constraints is fixed. In this case, we must first eliminate
$S_{i}^{a}$ and $T_{i}^{a}$; otherwise we might conclude that $\pi^{Ua}$
is second class as $\left\lbrace \pi^{Ua},T_{j}^{b}\right\rbrace \neq0$.

When quantizing the model of eq. (1) using the path integral, eqs.
(13,14) show that if second-class constraints are present, the factor
of $\Theta=\det^{1/2}\left\lbrace \theta_{a},\theta_{b}\right\rbrace $
contributes to the measure. This can be exponentiated by use of a
Grassmann ``ghost'' field $d^{a}$ and hence absorbed into the effective
action, %eq38
\[
\Theta=\int Dd^{a}\exp i\int d^{4}x\, d^{a}\left\lbrace \theta_{a},\theta_{b}\right\rbrace d^{b}\eqno(38)
\]
\[
\hspace{-0.5cm}=\int Dd^{a}\exp i\int d^{4}x\,\mathcal{L}_{ghost}.
\]
Unlike the massive vector Proca model of eq. (15), this factor of
$\Theta$ for the model of eq. (1) is not constant, but rather is
field-dependent as can be seen from eq. (37). In fact, eq. (35a) shows
that for this model %eq39
\[
\Theta=\det\left\lbrace S_{i}^{a},T_{j}^{b}\right\rbrace \eqno(39)
\]
which by eqs. (37,38) becomes %eq
\[
=\int D\overline{d}^{a}Dd^{a}exp\, i\int d^{4}x\mathcal{L}_{ghost}
\]
where %eq40
\[
\mathcal{L}_{ghost}=\overline{d}^{a}\left\lbrace S_{i}^{a},T_{\ell}^{x}\right\rbrace d_{\ell}^{x}\eqno(40)
\]
where $\left\lbrace S_{i}^{a},T_{\ell}^{x}\right\rbrace $ is given
by eq. (37b). Unfortunately, the calculations of $<WW>$, $<W\phi>$
and $<\phi\phi>$ in refs. {[}2-4{]} are deficient as the contribution
coming from eq. (40) has been ignored.

One might try to circumvent having to include the contribution to
radiative corrections coming from the field dependent of $\Theta$
by introducing a Stueckelberg field to restore gauge invariance, as
was done in the case of the Proca model discussed above. This would
hopefully eliminate all second-class constraints in the theory, making
it feasible to employ the FP quantization procedure. Unfortunately,
not all second-class constraints are eliminated upon introduction
of a Stueckelberg field and so this hope is not realized.

To show this, we begin by introducing a Stueckelberg vector field
$\sigma_{\mu}^{a}$. The most straightforward way of doing this is
to replace $\phi_{\mu\nu}^{a}$ by $\phi_{\mu\nu}^{a}+D_{\mu}^{ab}\sigma_{\nu}^{b}-D_{\nu}^{ab}\sigma_{\mu}^{b}$
in eq. (1). This leads to %eq41
\[
\mathcal{L}=-\frac{1}{4}F_{\mu\nu}^{a}F^{a\mu\nu}+\frac{1}{12}\left[G_{\mu\nu\lambda}^{a}+f^{abc}\left(F_{\mu\nu}^{b}\sigma_{\lambda}^{c}+F_{\nu\lambda}^{b}\sigma_{\mu}^{c}+F_{\lambda\mu}^{b}\sigma_{\nu}^{c}\right)\right]
\]
\[
\left[G^{a\mu\nu\lambda}+f^{abc}\left(F^{b\mu\nu}\sigma^{c\lambda}+F^{b\nu\lambda}\sigma^{c\mu}+F^{b\lambda\mu}\sigma^{c\nu}\right)\right]
\]
\[
+\frac{m}{4}\epsilon^{\mu\nu\lambda\sigma}\phi_{\mu\nu}^{a}F_{\lambda\sigma}^{b}+\frac{\mu^{2}}{8}\epsilon^{\mu\nu\lambda\sigma}\left(\phi_{\mu\nu}^{a}\phi_{\lambda\sigma}^{a}+4\phi_{\mu\nu}^{a}D_{\lambda}^{ab}\sigma_{\sigma}^{b}\right.
\]
\[
\left.-2f^{abc}F_{\mu\nu}^{a}\sigma_{\lambda}^{b}\sigma_{\sigma}^{c}\right).\eqno(41)
\]
By construction, this Lagrangian is invariant under not only the gauge
transformation of eqs. (5,6), but also the gauge transformation %eq42
\[
\delta\phi_{\mu\nu}^{a}=D_{\mu}^{ab}w_{\nu}^{b}-D_{\nu}^{ab}w_{\mu}^{b}
\]
\[
\hspace{-2.1cm}\delta\sigma_{\mu}^{a}=-w_{\mu}^{a}\eqno(42)
\]
\[
\hspace{-2.7cm}\delta W_{\mu}^{a}=0.
\]

Despite the presence of this new gauge invariance we unfortunately
still have second-class constraints in the model. To see this, we
perform a constraint analysis of the model of eq. (41). It is sufficient
to establish these points to set $W_{\mu}^{a}=m=0$ and to look at
the Abelian limit.

In this case, we begin by defining %eq43
\[
A_{i}=\phi_{0i},\quad B_{i}=\frac{1}{2}\epsilon_{ijk}\phi_{jk},\quad S=\sigma_{0},\quad R_{i}=\sigma_{i}\eqno(43a-d)
\]
so that from eq. (41) %eq44
\[
\hspace{-1.4cm}\mathcal{L}=\frac{1}{2}\dot{B}_{k}\dot{B}_{k}-\dot{B}_{i}\epsilon_{ijk}\partial_{j}A_{k}+\frac{1}{2}\left(\partial_{k}A_{m}\partial_{k}A_{m}-\partial_{k}A_{m}\partial_{m}A_{k}\right)\eqno(44)
\]
\[
-\frac{1}{2}(\partial_{k}B_{k})^{2}+\mu^{2}\left(A_{k}B_{k}+\epsilon_{ijk}A_{i}\partial_{j}R_{k}+B_{i}\dot{R}_{i}-B_{i}\partial_{i}S\right).
\]
The canonical momenta associated with these fields are respectively
%eq45a-d
\[
\pi_{i}^{A}=0,\quad\pi_{i}^{B}=\dot{B}_{i}-\epsilon_{ijk}\partial_{j}A_{k},\quad\pi^{S}=0,\quad\pi_{i}^{R}=\mu^{2}B_{i}\eqno(45a-d)
\]
and the Hamiltonian is %eq46
\[
\mathcal{H}=\frac{1}{2}\pi_{k}^{B}\pi_{k}^{B}+A_{i}\epsilon_{ijk}\partial_{j}(\pi_{k}^{B}-\mu^{2}R_{k})-\mu^{2}B_{k}(A_{k}-\partial_{k}S)+\frac{1}{2}(\partial_{i}B_{i})^{2}.\eqno(46)
\]
The primary constraints $(\pi_{i}^{A},\pi^{S},\pi_{i}^{R}-\mu^{2}B_{i})$
yield respectively the secondary constraints %eq47a-c
\[
B_{i}+\epsilon_{ijk}\left(\partial_{j}R_{k}-\frac{1}{\mu^{2}}\partial_{j}\pi_{k}^{B}\right)=0\eqno(47a-c)
\]
\[
\mu^{2}\partial_{i}B_{i}=0
\]
\[
-\mu^{2}\pi_{i}^{B}=0.
\]
In turn, we see that since %eq48a-b
\[
\hspace{-4.1cm}\left\lbrace \partial_{i}B_{i},\mathcal{H}\right\rbrace =\partial_{i}\pi_{i}^{B}\eqno(48a-b)
\]
\[
\left\lbrace \pi_{i}^{B},\mathcal{H}\right\rbrace =\partial_{i}\partial_{j}B_{j}+\mu^{2}A_{i}-\partial_{i}S
\]
there are no tertiary constraints. In total, when $\mu^{2}\neq0$,
there are five first-class constraints ($\pi_{k}^{A}$, $\pi^{S}$
and $\pi_{i}^{RL}$) and eight second-class constraints ($\pi_{i}^{B}$,
$B_{i}^{L},(\pi_{i}^{R}-\mu^{2}B_{i})^{T}$ and $\epsilon_{ijk}\partial_{j}R_{k}^{T}+B_{i}^{T}$)
where $L$ and $T$ refer to the longitudinal and transverse component
of a vector in three dimensions. When we also include the five gauge
conditions associated with the first-class constraints, we see that
there are 18 constraints on the 20 variables ($\phi_{\mu\nu}$, $\sigma_{\mu}$
and their associated momenta) in phase space leaving us just two with
physical degrees of freedom upon having introduced the field $\sigma_{\mu}$
in addition to $\phi_{\mu\nu}$. These degrees of freedom decouple
from the tensor field, much as the Stueckelberg field in the Proca
model (eq. (22)) decouples from the vector field. To see this, in
$\mathcal{L}$ given by eq. (44), the shift $A_{i}\rightarrow A_{i}-\dot{R}_{i}^{L}+\partial_{i}S$
eliminates $-\dot{R}_{i}^{L}+\partial_{i}S$ and its associated momentum
from the model.

Unlike the case of the Proca field, we see that introduction of a
Stueckelberg field to restore a gauge invariance absent in the original
Lagrangian does not eliminate all second-class constraints in the
model (There may be other, less obvious ways of introducing a Stueckelberg
field that eliminates all second class constraints). From our discussion
of the constraints arising from the action of eq. (44), it is readily
apparent that the full theory of eq. (41) has second-class constraints
whose PB involves the fields $\phi_{\mu\nu}$ and $\sigma_{\mu}$.
Consequently the contribution of the factor $\Theta$ to the measure
of the path integral of eq. (13) will again be non-trivial even though
the gauge invariance of eq. (42) is now present.

One might also attempt to eliminate the problems associated with incorporating
$\Theta$ into the measure of the path integral by either directly
converting second-class constraints into first-class ones {[}34,35{]},
by treating half of the second-class constraints as begin first-class
and the other half as being associated gauge conditions {[}36,37{]}
or by introduction of new variables to convert second-class constraints
into first-class ones {[}25-28{]}. (This is the BFT approach discussed
above.) It doesn't appear to be feasible to employ any of these three
approaches to eliminate the second-class constraints present in the
model of eq. (1) if both the constraints and the Hamiltonian are to
be in closed form. However, in an appendix we will pursue the third
approach (BFT) in the limit in which the gauge field is eliminated
from this model and there is but a single tensor field $\phi_{\mu\nu}$,
and to examine the feasibility of converting half of the second class
constraints to gauge conditions as in refs. {[}36,37{]}.

\section{Conversion to the Lagrangian in the Path Integral}

The path integral of eq. (13) is not in manifestly covariant form.
For many models though, once the integral over $p_{i}(t)$ is performed,
the phase of the exponential is given by the action \\
$S=\int dt(L(q^{i}(t),\dot{q}^{i}(t))$ and manifest covariance is
restored. This is immediately true in a scalar theory with quartic
self interaction. It is also true in YM theory as there the first-class
constraints that are present make a contribution to the measure of
eq. (14) that is equivalent to that which arises in the manifestly
covariant FP procedure in the Lorenz-Feynman gauge.

However, this equivalence is not necessarily true in all gauge theories.
The problem of reconciling the path integral of eqs. (13,14) which
arises out of canonical quantization with the path integral derived
by factoring out the integral over gauge equivalent field configuration
with the phase factor taken to be $\exp\, iS$ {[}16,17{]} has been
considered in {[}16,42{]} in the context of the action being the second
order Einstein-Hilbert action. We wish to investigate more closely
if the path integral of eqs. (13,14) when applied to the model of
eq. (1) (with $m=0$) is equivalent to the FP path integral used for
this model in refs. {[}2-4{]}. The approach of Garczynski {[}24{]}
will be used in this discussion.

In order to effect an integration over $p_{i}(t)$ in eq. (13), we
begin by noting that in a system with no constraints and $n$ degrees
of freedom %eq49
\[
p_{i}=\frac{\partial L(q^{i},\dot{q}^{i})}{\partial\dot{q}^{i}}\eqno(49)
\]
and so, using the standard property of the Dirac delta function %eq50
\[
dx\,\delta(f(x))=dx\sum_{i}\delta(x-a_{i})/|f^{\prime}(a_{i})|\quad(f(a_{i})=0)\eqno(50)
\]
we see that %eq51
\[
\prod_{k=1}^{n}dv^{k}\delta(v^{k}-\dot{q}^{k}(q^{k},v))=|A_{n}(q,v)|\prod_{k=1}^{n}dv^{k}\delta\left(\frac{\partial L(q,v)}{\partial v^{k}}-p_{k}\right).\eqno(51)
\]
where %eq52
\[
A_{n}(q,\dot{q})=\det\left(\frac{\partial^{2}L(q,\dot{q})}{\partial\dot{q}^{i}\partial\dot{q}^{j}}\right)\eqno(52)
\]
is the Hessian for the system. In this case, eq. (13) becomes %eq53
\[
<\mathrm{out}|S|\mathrm{in}>=\int\exp i\int_{-\infty}^{\infty}\left[p_{i}(\dot{q}^{i}-v^{i})+L(q^{i},v^{i})\right]\delta(v^{i}-\dot{q}^{i}(q,p))
\]
\[
Dq^{i}Dp_{i}D\, v^{i}\;.\eqno(53)
\]
The integral over $p_{i}$ can be done in eq. (53) upon using eq.
(51), leaving us with %eq54
\[
=\int|A_{n}(q,v)|\exp i\int_{-\infty}^{\infty}\left[\frac{\partial L(q,v)}{\partial v^{i}}(\dot{q}^{i}-v^{i})+L(q^{i},v^{i})\right]Dq^{i}Dv^{i}\;.\eqno(54)
\]
If now we let $v^{i}\rightarrow v^{i}+\dot{q}^{i}$, eq. (54) leads
to %eq55
\[
<\mathrm{out}|S|\mathrm{in}>=\int\exp i\int_{-\infty}^{\infty}dt\left[L(q^{i},\dot{q}^{i})\right]m(q^{i},\dot{q}^{i})Dq^{i}.\eqno(55)
\]
where %eq56
\[
m(q^{i},\dot{q}^{i})=\int|A_{n}(q^{i},\dot{q}^{i}+v)|\exp i\int_{-\infty}^{\infty}dt\left[L(q,\dot{q}+v)\right.\eqno(56)
\]
\[
\left.\quad-L(q,\dot{q})-v^{i}\frac{\partial}{\partial v^{i}}L(q,\dot{q}+v)\right]Dv^{i}.
\]
This approach to the path integral of eq. (13) can be adapted quite
easily to the case in which constraints are present. In the presence
of constraints, the Hessian $A_{n}(q,\dot{q})$ vanishes and the matrix
$\partial^{2}L(q,\dot{q})/\partial\dot{q}^{i}\partial\dot{q}^{j}$
has rank $r<n$. If eq. (49) can be solved for $i=1\ldots r$ then
we have %eq57
\[
\dot{q}^{i}=f^{i}(q^{1}\ldots q^{n},\, p_{1}\ldots p_{r},\,\dot{q}^{r+1}\ldots\dot{q}^{n})\,\,(i=1\ldots r).\eqno(57)
\]
We denote the first $r$ variables with a prime $(q^{i\prime},p_{i}^{\prime};\, i=1\ldots r)$
and all others with a double prime $(q^{i\prime\prime},p_{i}^{\prime\prime};\, i=r+1\ldots n)$.
Following ref. {[}24{]}, we find that the general path integral of
eq. (13) with constraints present leads to eq. (55) with $m(q,\dot{q})$
in eq. (56) now given by %eq58
\[
m(q^{i},\dot{q}^{i})=\int\left\lbrace |A_{r}(q,\dot{q}^{\prime}+v^{\prime},\dot{q}^{\prime\prime})|\exp i\int_{-\infty}^{\infty}dt\big[L(q,\dot{q}^{\prime}+v^{\prime},\dot{q}^{\prime\prime})\right.
\]
\[
-L(q,\dot{q})-v^{\prime i}\frac{\partial}{\partial v^{\prime i}}L(q,\dot{q}^{\prime}+v^{\prime},\dot{q}^{\prime\prime}\big]\eqno(58)
\]
\[
\left.S\big[q^{i},\frac{\partial}{\partial v^{\prime i}}L(q,\dot{q}^{\prime}+v^{\prime},\dot{q}^{\prime\prime}),v^{\prime\prime}\big]\right\rbrace Dv^{\prime i}Dv^{\prime\prime i}.
\]
In eq. (58), S is given by %eq59
\[
S[q^{i},p_{j}]=\det{^{1/2}}\left\lbrace \theta_{a},\theta_{b}\right\rbrace \det\left\lbrace \phi_{a},\chi_{b}\right\rbrace \delta(\theta_{a})\delta(\phi_{a})\delta(\chi_{a})\eqno(59)
\]
and %eq60
\[
A_{r}(q,\dot{q}^{\prime},\dot{q}^{\prime\prime})=\det\left(\frac{\partial^{2}L(q,\dot{q}^{\prime},\dot{q}^{\prime\prime})}{\partial\dot{q}^{\prime i}\partial\dot{q}^{\prime j}}\right).\eqno(60)
\]

We now can apply the path integral of eqs. (55) and (58) to the model
of eq. (1). (We set $m=0$ for purposes of illustration.) The Lagrangian
$L$ is clearly manifestly covariant. We then identify $q^{\prime i}$
and $q^{\prime\prime i}$ with $(B_{i}^{a},V_{i}^{a})$ and $(A_{i}^{a},U^{a})$
respectively, and take $v^{\prime i}$ and $v^{\prime\prime i}$ to
be $(\beta_{i}^{a},\nu_{i}^{a})$ and $(\alpha_{i}^{a},\mu_{i}^{a})$
respectively. This leads to the argument of the exponential in eq.
(58) to being %eq61
\[
\int dt\left[L(q,\dot{q}^{\prime}+v,\dot{q}^{\prime\prime})-L(q,\dot{q})-v^{\prime i}\frac{\partial}{\partial v^{\prime i}}L(q,\dot{q}^{\prime}+v^{\prime},\dot{q}^{\prime\prime})\right]
\]
\[
=-\frac{1}{2}\int d^{4}x\left[(\nu_{i}^{a})^{2}+(\beta_{i}^{a})^{2}\right].\eqno(61)
\]
The contribution of $A_{r}$ to eq. (58) is just a constant and consequently
can be factored out of the path integral. However, from eq. (59) we
see that $S$ is non-trivial. Upon choosing the gauge conditions %eq62abc
\[
U^{a}=0\eqno(62a)
\]
\[
A_{i}^{a}=0\eqno(62b)
\]
\[
\partial_{i}V_{i}^{a}=0\eqno(62c)
\]
to be associated with the first-class constraints of eqs. (32a,c;
34a) respectively, we find that %eq63
\[
\det\left\lbrace \phi_{a},\chi_{b}\right\rbrace =\det\left[\partial_{i}D_{i}^{ab}\right].\eqno(63)
\]
We now can use eq. (37) to obtain the contributions of the second
class constraints to $S$.

In pure YM theory, one only has first-class constraints which lead
to %eq64
\[
S=\det(\partial_{i}D_{i}^{ab})\delta(\pi^{Ua})\delta(U^{a})\delta(\partial_{i}V_{i}^{a})\delta(D_{i}^{ab}\pi_{u}^{Vb})\eqno(64)
\]
for $S$. As was shown in ref. {[}20{]} (see also ref. {[}43{]}),
this can be replaced by the FP factor in the Lorenz-Feynman gauge
($\det(\partial^{\mu}D_{\mu}^{ab})\delta(\partial^{\mu}W_{\mu}^{a})$),
which ensures that the path integral expression for $<\mathrm{out}|S|\mathrm{in}>$
is manifestly covariant.

In contrast, the contribution from $S$ to the measure $m$ of eq.
(58) for the tensor-vector model receives the contribution %eq65
\[
\hspace{-3cm}S\left(q^{i},\frac{\partial}{\partial v^{\prime i}}L(q,\dot{q}^{\prime}+v^{\prime},v^{\prime\prime})\right)\eqno(65)
\]
\[
=\left[\delta(U^{a})\delta(A_{i}^{a})\delta(\partial_{i}V_{i}^{a})\right]\bigg[\delta(\mu^{a})\delta(\alpha_{i}^{a})
\]
\[
\delta\left(D_{i}^{ab}(\dot{V}+mB+\nu)_{i}^{b}+f^{abc}B_{i}^{b}(\dot{B}+\beta)_{i}^{c}\right)\bigg]
\]
\[
\bigg[\delta\left(\epsilon_{ijk}D_{j}^{ab}(\dot{B}+\beta)_{k}^{b}-\mu^{2}B_{i}^{a}-\frac{m}{2}\epsilon_{ijk}F_{jk}^{a}\right)\bigg]
\]
\[
\bigg[\delta\bigg(-\mu^{2}(\dot{B}+\beta)_{i}^{a}+\epsilon_{ijk}\bigg(f^{abc}\left((\dot{V}+mB+\nu)_{j}^{b}(\dot{B}+\beta)_{k}^{c}-mB_{j}^{b}(\dot{B}+\beta)_{k}^{c}\right)
\]
\[
+(D_{j}D_{\ell}D_{k}D_{\ell})^{a}\bigg)\bigg)\bigg]\left[\det(\partial_{i}D_{i}^{ab})\right]
\]
\[
det^{1/2}\left[f^{apm}f^{xpn}\left(\delta_{i\ell}(\dot{B}+\beta)_{k}^{m}(\dot{B}+\beta)_{k}^{n}-(\dot{B}+\beta)_{\ell}^{m}(\dot{B}+\beta)_{\ell}^{n}\right)\right.
\]
\[
+\delta_{i\ell}(D_{j}D_{k}D_{k}D_{j}-D_{j}D_{j}D_{k}D_{k})^{ax}
\]
\[
+(D_{\ell}D_{i}D_{j}D_{j}+D_{j}D_{j}D_{\ell}D_{i}-D_{j}D_{i}D_{\ell}D_{j}-D_{\ell}D_{j}D_{j}D_{i})^{ax}
\]
\[
\left.+\mu^{2}\bigg(\mu^{2}\delta^{ax}\delta_{i\ell}-\epsilon_{i\ell m}f^{axp}\bigg((\dot{V}+mB+\nu)_{m}^{p}+mB_{m}^{p}\bigg)\bigg)\right].
\]
The terms in square brackets on the right side of eq. (65) come in
turn from the gauge conditions (eq. (62)), the first class constraints
(eqs. (32a,c; 34a)), the second class constraints (eqs. (34b, 36)),
the PB of the first class constraints with the gauge conditions (eq.
(63)), and the PB of the second class constraints (eq. (37b)) respectively.
We have used the fact the
\[
\frac{\partial\mathcal{L}}{\partial\dot{B}_{i}^{a}}=\dot{B}_{i}^{a}+f^{abc}U^{b}B_{i}^{c}-\epsilon_{ijk}(D_{j}A_{k})^{a}\eqno(66a)
\]
\[
\frac{\partial\mathcal{L}}{\partial\dot{V}_{i}^{a}}=\dot{V}_{i}^{a}-D_{i}^{ab}U^{b}+mB_{i}^{a}.\eqno(66b)
\]
When eq. (65) is combined with eq. (61), it is apparent that for this
tensor-vector model, the contribution of $m$ to the path integral
in configuration space is both non-trivial and is not manifestly covariant.

\section{Discussion}

Finding a way of quantizing a model in a way that is manifestly consistent
with covariance has been a long standing problem, especially since
the procedure in which a classical PB is converted into a quantum
commutator and time evolution is governed by a Hamiltonian is always
specific to one preferred reference frame. Stueckelberg {[}44{]} was
the first to employ a manifestly covariant approach to quantizing
electrodynamics; this was followed by the work of Feynman {[}45{]},
Schwinger {[}46{]} and Tomonaga {[}47{]}. The path integral has long
been seen as a way of quantizing any gauge model in a way consistent
with manifest covariance. (It is not readily apparent if this is true
for theories other than Yang-Mills theory which contain only first
class constraints.) The original work of Faddeev {[}20{]} for a restricted
class of first-class constraints has been subsequently extended {[}42,
48-50, 7-13{]} to show how this can be done. However, incorporation
second-class constraints (which lead to the factor of $\det^{1/2}\left\lbrace \theta_{a},\theta_{b}\right\rbrace $
in eqs. (14) and (59)) into the path inegral in a way that is manifestly
covariant has not been done, though general discussions involving
second-class constraints in the context of the path integral have
appeared in the literature {[}51-54{]}. The model of eq. (1) provides
for the first time an example where this particular problem becomes
acute, as can be seen from eq. (64). We have examined the possibility
of converting second-class constraints to being first-class through
introduction of ``Stueckelberg fields'', thereby making it feasible
to employ techniques developed for models containing only first-class
constraints, but unlike the Proca model, this is not possible for
the non-Abelian model of eq. (1) as second-class constraints remain
even after introduction of Stueckelberg fields. Consequently, this
approach is distinct from the BFT approach in which all second-class
constraints became first-class through introduction of new auxiliary
fields. Furthermore, as shown in the appendix, it seems to be impossible
in our non-Abelian model to devise a simple way of modifying the Hamiltonian
so that our model can be viewed as merely being the gauge-fixed limit
of a manifestly covariant model with only first-class constraints.
This latter approach to handling second hand constraints has been
employed in conjunction with BRST quantization {[}55{]}. The question
of including secondary second class constraints into the path integral
involving the exponential of the Lagrangian has also been examined
in ref. {[}56{]}.

The model of eq. (1) which we have been dealing with in this paper
is clearly of limited physical interest. However, the problem of properly
quantizing this model is highly non-trivial and provides insight that
likely will give an understanding of how more significant theories
should be properly quantized. It is apparent that one such theory
is gravity, as is described by the Einstein-Hilbert action. Second-class
constraints occur in the first-order (``Palatini'') formulation
of the Einstein-Hilbert action {[}40,41{]}. From the results of this
paper, we see that it is necessary to take into account the contribution
of ghost loops arising from these second-class constraints when computing
radiative affects using this first-order action. (The Faddeev-Popov
procedure is inadequate to quantize this model.) The Palatini action
is worth analyzing as it is just cubic in the interaction terms {[}57,58{]},
while the second-order form of the Einstein-Hilbert action is non-polynomial
{[}38,39{]}. We are examining this issue now. We also note the presence
of second class constraints whose PB is non-trivial in the first order
formulation of the three-dimensional Einstein-Cartan action {[}59{]}.

The question of how to convert the path integral from an integral
in phase space to an integral in configuration space is also important
when quantizing the Palatini action for general relativity\vspace{0.2cm}
. \\
 \textbf{\Large {Acknowledgments}}\textbf{}\\
\textbf{ Roger Macleod had a helpful suggestion.}

\noindent \newpage{}

\noindent \textbf{\Large {Appendix}}\textbf{\vspace{1cm}
}\\
\textbf{ In this appendix, we will first discuss how the BFT approach
of refs. {[}25-28{]} can be applied to a limiting case of the model
of eq. (1) in which %eqA1
\[
\mathcal{L}=\frac{1}{12}\left(\partial_{\mu}\phi_{\nu\lambda}+\partial_{\nu}\phi_{\lambda\mu}+\partial_{\lambda}\phi_{\mu\nu}\right)\left(\partial^{\mu}\phi^{\nu\lambda}+\partial^{\nu}\phi^{\mu\lambda}+\partial^{\lambda}\phi^{\mu\nu}\right)\eqno(A.1)
\]
\[
+\frac{\mu^{2}}{8}\epsilon^{\mu\nu\lambda\phi}\phi_{\mu\nu}\phi_{\lambda\sigma}.
\]
(It is trivial to also include {[}1{]} the terms $-\frac{1}{4}(\partial_{\mu}W_{\nu}-\partial_{\nu}W_{\mu})^{2}+\frac{m}{4}\epsilon_{\mu\nu\lambda\sigma}\phi^{\mu\nu}(\partial^{\lambda}W^{\sigma}-\partial^{\sigma}W^{\lambda})$
into $\mathcal{L}$ in the following discussion.) Using the definitions
%eqA2
\[
A_{i}=\phi_{0i}\qquad B_{i}=\frac{1}{2}\epsilon_{ijk}\phi_{jk}\eqno(A.2a,b)
\]
it follows that %eqA3
\[
\mathcal{L}=\frac{1}{2}(\dot{B}_{i}-\epsilon_{ijk}\partial_{j}A_{k})^{2}-\frac{1}{2}(\partial_{i}B_{i})^{2}+\mu^{2}A_{i}B_{i}\,.\eqno(A.3)
\]
The momenta corresponding to $A_{i}$ and $B_{i}$ are respectively
%eqA4
\[
\pi_{i}^{A}=0,\qquad\pi_{i}^{B}=\dot{B}_{i}-\epsilon_{ijk}\partial_{j}A_{k}.\eqno(A.4a,b)
\]
The Hamiltonian is consequently %eqA5
\[
\hspace{-4.5cm}\mathcal{H}=\pi_{i}^{A}\dot{A}_{i}+\pi_{i}^{B}\dot{B}_{i}-\mathcal{L}
\]
\[
=\frac{1}{2}\pi_{i}^{B}\pi_{i}^{B}+\pi_{i}^{B}\epsilon_{ijk}\partial_{j}A_{k}+\frac{1}{2}(\partial_{i}B_{i})^{2}-\mu^{2}A_{i}B_{i}\eqno(A.5)
\]
and it follows that the primary constraint of eq. (A.4a) leads to
the secondary constraint %eqA6
\[
S_{i}=\epsilon_{ijk}\partial_{j}\pi_{k}^{B}-\mu^{2}B_{i}\eqno(A.6)
\]
and subsequently to the tertiary constraint %A7
\[
T_{i}=\pi_{i}^{B}\,.\eqno(A.7)
\]
It is apparent that $\pi_{i}^{A}=0$ is a first-class constraint and
$S_{i}=T_{i}=0$ are second-class constraints as $\left\lbrace S_{i},T_{j}\right\rbrace =-\mu^{2}\delta_{ij}$.}

In keeping with the BFT approach, auxiliary fields are introduced
to convert the second-class constraints to first-class ones. Calling
these fields $Q_{i}$ and $P_{i}$, with %eqA8
\[
\left\lbrace Q_{i},P_{j}\right\rbrace =\delta_{ij}\eqno(A.8)
\]
we now form %eqA9ab
\[
\overline{S}_{i}=S_{i}+\mu^{2}Q_{i},\qquad\overline{T}_{i}=T_{i}+P_{i}\eqno(A.9ab)
\]
and %eqA10
\[
\overline{\mathcal{H}}=\mathcal{H}+(\partial_{i}\partial_{j}B_{j}+\mu^{2}A_{i})Q_{i}+2\pi_{i}^{B}P_{i}\eqno(A.10)
\]
\[
-\frac{1}{2}Q_{i}\partial_{i}\partial_{j}Q_{j}+\frac{3}{2}P_{i}P_{i}\,.
\]
These new quantities satisfy %eqA11a-c
\[
\left\lbrace \overline{S}_{i},\overline{T}_{j}\right\rbrace =0,\qquad\left\lbrace \overline{S}_{i},\overline{\mathcal{H}}\right\rbrace =\mu^{2}\overline{T}_{i},\qquad\left\lbrace \overline{T}_{i},\overline{\mathcal{H}}\right\rbrace =0\eqno(A.11a-c)
\]
so that in this new ``bared'' system $(\pi_{i}^{A},\overline{S}_{i},\overline{T}_{i})$
are all first class constraints when considered in conjunction with
the Hamiltonian $\overline{\mathcal{H}}$ with $(\overline{S}_{i},\overline{T}_{i},\overline{\mathcal{H}})$
reducing to $(S_{i},T_{i},\mathcal{H})$ as $(Q_{i},P_{i})$ go to
zero. This may be regarded as making a particular choice of gauge.

The Lagrangian in the ``bared'' system can be found by considering
%eqA12
\[
\overline{\mathcal{L}}=\pi_{i}^{A}\dot{A}_{i}+\pi_{i}^{B}\dot{B}_{i}+P_{i}\dot{Q}_{i}-\overline{H}\eqno(A.12)
\]
and using the equations of motion for $\dot{A}_{i}$, $\dot{B}_{i}$,
$\dot{Q_{i}}$ to eliminate the momenta $\pi_{i}^{A}$, $\pi_{i}^{B}$
and $P_{i}$ %eqA13abc
\[
\hspace{-3.5cm}\dot{A}_{i}=0\eqno(A.13a)
\]
\[
P_{i}=2(\dot{B}_{i}-\epsilon_{ijk}\partial_{j}A_{k})-\dot{Q}_{i}\eqno(A.13b)
\]
\[
\pi_{i}^{B}=2\dot{Q}_{i}-3(\dot{B}_{i}-\epsilon_{ijk}\partial_{j}A_{k})\eqno(A.13c)
\]
to express $\overline{\mathcal{L}}$ in terms of the ``position''
variables $(A_{i},B_{i},Q_{i})$ and their associated velocities.
We obtain %eqA14
\[
\overline{\mathcal{L}}=-\frac{1}{2}\dot{Q}_{i}\dot{Q}_{i}-\frac{3}{2}(\dot{B}_{i}-\epsilon_{ijk}\partial_{j}A_{k})^{2}+2(\dot{B}_{i}-\epsilon_{ijk}\partial_{j}A_{k})\dot{Q}_{i}
\]
\[
\hspace{2.1cm}-\frac{1}{2}(\partial_{i}B_{i})^{2}+\mu^{2}A_{i}B_{i}-(\partial_{i}\partial_{j}B_{j}+\mu^{2}A_{i})Q_{i}+\frac{1}{2}Q_{i}\partial_{i}\partial_{j}Q_{j}\,.\eqno(A.14)
\]
Unlike the case considered in ref. {[}29,30{]}, the gauge choice in
which $Q_{i}=0$ does not reduce $\overline{\mathcal{L}}$ in eq.
(A.14) to $\mathcal{L}$ in eq. (A.3). It does not appear to be possible
to express $\overline{\mathcal{L}}$ in a manifestly covariant form.

To find the gauge transformation generated by the first class constraints
$\gamma_{i}^{(N)}=(\pi_{i}^{A},\overline{S}_{i},\overline{T}_{i})$
we use the HTZ method {[}32{]}. (One could also employ the technique
of ref. {[}31{]}.) We begin by introducing a gauge generator %eq15
\[
G=\lambda_{i}^{(1)}\pi_{i}^{A}+\lambda_{i}^{(2)}\overline{S}_{i}+\lambda_{i}^{(3)}\overline{T}_{i}\eqno(A.15)
\]
and then employing the equation {[}32{]} %eqA16
\[
\frac{D\lambda_{i}^{(N)}}{Dt}\gamma_{i}^{(N)}+\left\lbrace \lambda_{i}^{(N)}\gamma_{i}^{(N)},\,\overline{\mathcal{H}}+U_{i}^{(1)}\gamma_{i}^{(1)}\right\rbrace -\delta U_{i}^{(1)}\gamma_{i}^{(1)}=0\eqno(A.16)
\]
to derive the equations %eq17abc
\[
\dot{\lambda}_{i}^{(1)}=\delta U_{i}^{(1)}\eqno(A.17a)
\]
\[
\hspace{-0.5cm}\dot{\lambda}_{i}^{(2)}=\overline{\lambda}_{i}^{(1)}\eqno(A.17b)
\]
\[
\hspace{0.4cm}\dot{\lambda}_{i}^{(3)}=-\mu^{2}\lambda_{i}^{(2)}\,.\eqno(A.17c)
\]
If $\lambda_{i}^{(3)}\equiv\epsilon_{i}$, then eq. (A.17) leads to
the generator %eqA18
\[
G=-\frac{1}{\mu^{2}}\ddot{\epsilon}_{i}\pi_{i}^{A}-\frac{1}{\mu^{2}}\dot{\epsilon}_{i}(\epsilon_{ijk}\partial_{j}\pi_{k}^{B}-\mu^{2}B_{i}+\mu^{2}Q_{i})+\epsilon(\pi_{i}^{B}+P_{i}),\eqno(A.18)
\]
and so %eq19abc
\[
\delta A_{i}=\left\lbrace A_{i},G\right\rbrace =-\frac{1}{\mu^{2}}\ddot{\epsilon}_{i}\eqno(A.19a)
\]
\[
\hspace{-0.3cm}\delta B_{i}=-\frac{1}{\mu^{2}}\epsilon_{ijk}\partial_{j}\dot{\epsilon}_{k}+\epsilon_{i}\eqno(A.19b)
\]
\[
\hspace{-2.3cm}\delta Q_{i}=\epsilon_{i}\,.\eqno(A.19c)
\]
One can verify that $\overline{\mathcal{L}}$ in eq. (A.14) is invariant
under the transformation of eq. (A.19).

In the Abelian limit considered in eq. (A.1), it is possible to treat
$S_{i}=0$ as being a first class constraint and take $T_{i}=0$ as
being the associated gauge condition provided we use that Hamiltonian
\[
\mathcal{H}_{M}=\mathcal{H}-\frac{1}{2}T_{i}T_{i}\eqno(A.20)
\]
since $\left\lbrace S_{i},\mathcal{H}_{M}\right\rbrace =0$ and $\mathcal{H}_{M}\big|_{T_{i}=0}=\mathcal{H}$
{[}36,37{]}.

We now consider applying the techniques of refs. {[}36,37{]} to the
full Hamiltonian of eq. (33). Finding the associated Hamiltonian $\mathcal{H}_{M}$
in closed form is not possible, but one can determine $\mathcal{H}_{M}$
in a perturbative fashion. With $\mathcal{H}_{C}$, $S_{i}^{a}$ and
$T_{i}^{a}$ being given by eqs. (33,34b,36) respectively, we take
\[
\mathcal{H}_{M}=\mathcal{H}_{C}+\frac{1}{2}F_{i{_{1}}i{_{2}}}^{a{_{1}}a{_{2}}}T_{i{_{1}}}^{a{_{1}}}T_{i{_{2}}}^{a{_{2}}}+\frac{1}{3}F_{i{_{1}}i{_{2}}i{_{3}}}^{a{_{1}}a{_{2}}a{_{3}}}T_{i{_{1}}}^{a{_{1}}}T_{i{_{2}}}^{a{_{2}}}T_{i{_{3}}}^{a{_{3}}}+\ldots\eqno(A.21)
\]
(where $F_{i{_{1}}\ldots i{_{n}}}^{a{_{1}}\ldots a{_{n}}}$ are functions
of the canonical variables and are symmetric in each set of indices)
then the requirement
\[
\left\lbrace S_{i}^{a},\mathcal{H}_{M}\right\rbrace =0\eqno(A.22)
\]
leads to
\[
0=T_{i}^{a}+\left(F_{i{_{1}}i{_{2}}}^{a{_{1}}a{_{2}}}\Delta_{ii{_{1}}}^{aa{_{1}}}T_{i{_{2}}}^{a{_{2}}}+\frac{1}{2}\left\lbrace S_{i,}^{a}F_{i{_{1}}i{_{2}}}^{a{_{1}}a{_{2}}}\right\rbrace T_{i{_{1}}}^{a{_{1}}}T_{i{_{2}}}^{a{_{2}}}\right)
\]
\[
+F_{i{_{1}}i{_{2}}i{_{3}}}^{a{_{1}}a{_{2}}a{_{3}}}\Delta_{ii{_{1}}}^{aa{_{1}}}T_{i{_{2}}}^{a{_{2}}}T_{i{_{3}}}^{a{_{3}}}+\ldots\eqno(A.23)
\]
where $\Delta_{ij}^{ab}=\left\lbrace S_{i}^{a},T_{j}^{b}\right\rbrace $.
We are led to a set of nested equations whose solution is
\[
F_{i{_{1}}i{_{2}}}^{a{_{1}}a{_{2}}}=-\Delta_{\;\;\;\; i{_{1}}i{_{2}}}^{-1a{_{1}}a{_{2}}}\eqno(A.24)
\]
\[
F_{i{_{1}}i{_{2}}i{_{3}}}^{a{_{1}}a{_{2}}a{_{3}}}=-\frac{1}{6}\left[\Delta_{i{_{3}}i}^{-1a{_{3}}a}\left\lbrace S_{i}^{a},F_{i{_{1}}i{_{2}}}^{a{_{1}}a{_{2}}}\right\rbrace +\Delta_{i{_{2}}i}^{-1a{_{2}}a}\left\lbrace S_{i}^{a},F_{i{_{3}}i{_{1}}}^{a{_{3}}a{_{1}}}\right\rbrace +\Delta_{i{_{1}}i}^{-1a{_{1}}a}\left\lbrace S_{i}^{a},F_{i{_{2}}i{_{3}}}^{a{_{2}}a{_{3}}}\right\rbrace \right]
\]
\qquad{}\qquad{}etc.\\
These solutions are quite complicated as can be seen from the explicit
expression for $\Delta_{ij}^{ab}$ in eq. (37b).

In the gauge $T_{i}^{a}=0$, $\mathcal{H}_{M}$ of eq. (A.27) reduces
to $\mathcal{H}_{C}$ of eq. (33).

We finally note that we can accommodate second class constraints by
solving for the Lagrange multipliers with which they are associated
in the extended Hamiltonian $\mathcal{H}_{E}$. For example, $\mathcal{H}_{c}$
for the Proca model in eq. (18) is associated with
\[
\mathcal{H}_{E}=\mathcal{H}_{c}+\mu_{1}\pi^{0}+\mu_{2}(\partial_{i}\pi^{i}+m^{2}A_{0});\eqno(A.25)
\]
in order that $\left\lbrace \theta_{i},\int\mathcal{H}_{E}dx\right\rbrace =0$
$(i=1,2)$ we have
\[
\mu_{1}=\partial_{i}A_{i}\qquad\mu_{2}=(\partial_{i}\pi^{i}+m^{2}A_{0})/(2m^{2}).\eqno(A.26)
\]
With these expressions for $\mu_{1}$ and $\mu_{2}$ we find that
\[
\hspace{-2.8cm}\dot{A}_{0}=\left\lbrace A_{0},\int\mathcal{H}_{E}dx\right\rbrace =\partial_{i}A_{i}\eqno(A.27a)
\]
\[
\dot{A}_{i}=\left\lbrace A_{i},\int\mathcal{H}_{E}dx\right\rbrace =(\delta_{ij}-\partial_{i}\partial_{j}/m^{2})\pi_{j}.\eqno(A.27b)
\]
It does not appear to be possible to derive $\mathcal{H}_{E}$ from
a covariant Lagrangian.
\end{document}